\documentclass[aps,prb,twocolumn,showpacs,preprintnumbers,amsmath,amssymb,superscriptaddress]{revtex4}
\usepackage{mathptm}  
\usepackage{dcolumn}                    
\usepackage{bm}                        
\usepackage{graphicx}
\usepackage{times}

\newcommand{\ve}[1]{\boldsymbol{#1}}
\newcommand{\vp}{\ve{p}} 

\newcommand{\s}{\hat{\ve{s}}} 
\newcommand{\h}{\ve{h}} 
\newcommand{\bs}{\boldsymbol{\hat{\sigma}}}

\newcommand{\e}[1]{\mathrm{e}^{#1}}

\newcommand{\eg}{\textit{e.g. }}
\newcommand{\etal}{\emph{et al.}}
\def\i{\mathrm{i}}

\begin{document}
\title[Pure spin-current generated by reflection at a normal metal$\mid$2DEG interface]{Pure spin-current generated by reflection at a normal metal$\mid$2DEG interface}
\author{Jacob Linder}
\affiliation{Department of Physics, Norwegian University of
Science and Technology, N-7491 Trondheim, Norway}
\author{Takehito Yokoyama}
\affiliation{Department of Applied Physics, University of Tokyo, Tokyo 113-8656, Japan}
\author{Asle Sudb{\o}}
\affiliation{Department of Physics, Norwegian University of
Science and Technology, N-7491 Trondheim, Norway}

\date{Received \today}

\begin{abstract}
\noindent The concept of a spin-current is a useful tool in understanding spin-transport in hybrid systems, but its very definition is problematic in systems where spin-orbit coupling effects are strong. In the absence of spin-dependent scattering, the spin-current remains well-defined. We here propose a method for generating pure spin-currents in a normal metal where the spin-current consequently does not suffer from the aforementioned problems pertaining to its very definition or spin-relaxation processes. More specifically, we show how an unpolarized incident charge-current can induce a pure transverse spin-current by means of scattering at a normal metal$\mid$2DEG interface. This occurs for both Rashba and Dresselhaus spin-orbit coupling. An experimental setup for observation of this effect is proposed.
\end{abstract}

\maketitle

\section{Introduction}

The study of spin-transport in hybrid systems with magnetic elements is of crucial importance both in order to understand the basic physics of spin-transport and to find new functional devices emerging from fundamental research \cite{zutic_rmp_04,Ganichev,dyakonov_pla_71, hirsch_prl_99, murakami_science_03, sinova_prl_04, kato_science_04, wunderlich_prl_05}. In this context, the idea of a spin-current is a natural extension of the traditional charge-current, and is a heavily employed tool in the characterization of spin transport. The most straight-forward definition of a spin-current is, in analogy with the charge-current, simply the spin carried by a particle times its velocity. However, there are subtleties associated with this definition, in particular when spin-orbit coupling is present in the system \cite{shi_prl_06, chen_prb_09}. To illustrate 
this point, consider the general continuity equation for spin density $\boldsymbol{S}$:
\begin{align}
\partial_t \boldsymbol{S} + \nabla \cdot \boldsymbol{j}_S = \mathcal{T}.
\end{align}
Here, $\boldsymbol{j}_S$ is the spin-current whereas $\mathcal{T}$ represents a spin-sink/source term that causes $\boldsymbol{j}_S$ to be non-conserved. For instance, the effect of spin-transfer torque, where a spin-current is absorbed by a magnetic order parameter, may be incorporated into $\mathcal{T}$. The term $\mathcal{T}$ will in general be present in systems where the spin operator does not commute with the Hamiltonian. Now, the problem with the above equation is that one may absorb a portion, or in fact the entirety, of $\mathcal{T}$ into the definition of the spin-current by writing $\mathcal{T} = -\nabla\cdot \boldsymbol{P}$, which holds for systems where the average spin torque density vanishes in the bulk. The continuity equation then takes the form $\frac{\partial \boldsymbol{S}}{\partial t} + \nabla \cdot \boldsymbol{j}_S' = 0$, rendering the spin-current $\boldsymbol{j}_S' = \boldsymbol{j}_S + \boldsymbol{P}$ to be a conserved quantity, as opposed to $\boldsymbol{j}_S$. Therefore, there is an inherent ambiguity in the spin-current since one may define it in an arbitrary way by combining elements of $\boldsymbol{j}_S$ and $\mathcal{T}$. At the same time, it is clear that in a normal metallic region without any spin-sink/source term, the conventional definition of the spin-current serves well and is conserved. This fact will feature prominently below. 

\begin{figure}[b!]
\centering
\resizebox{0.45\textwidth}{!}{
\includegraphics{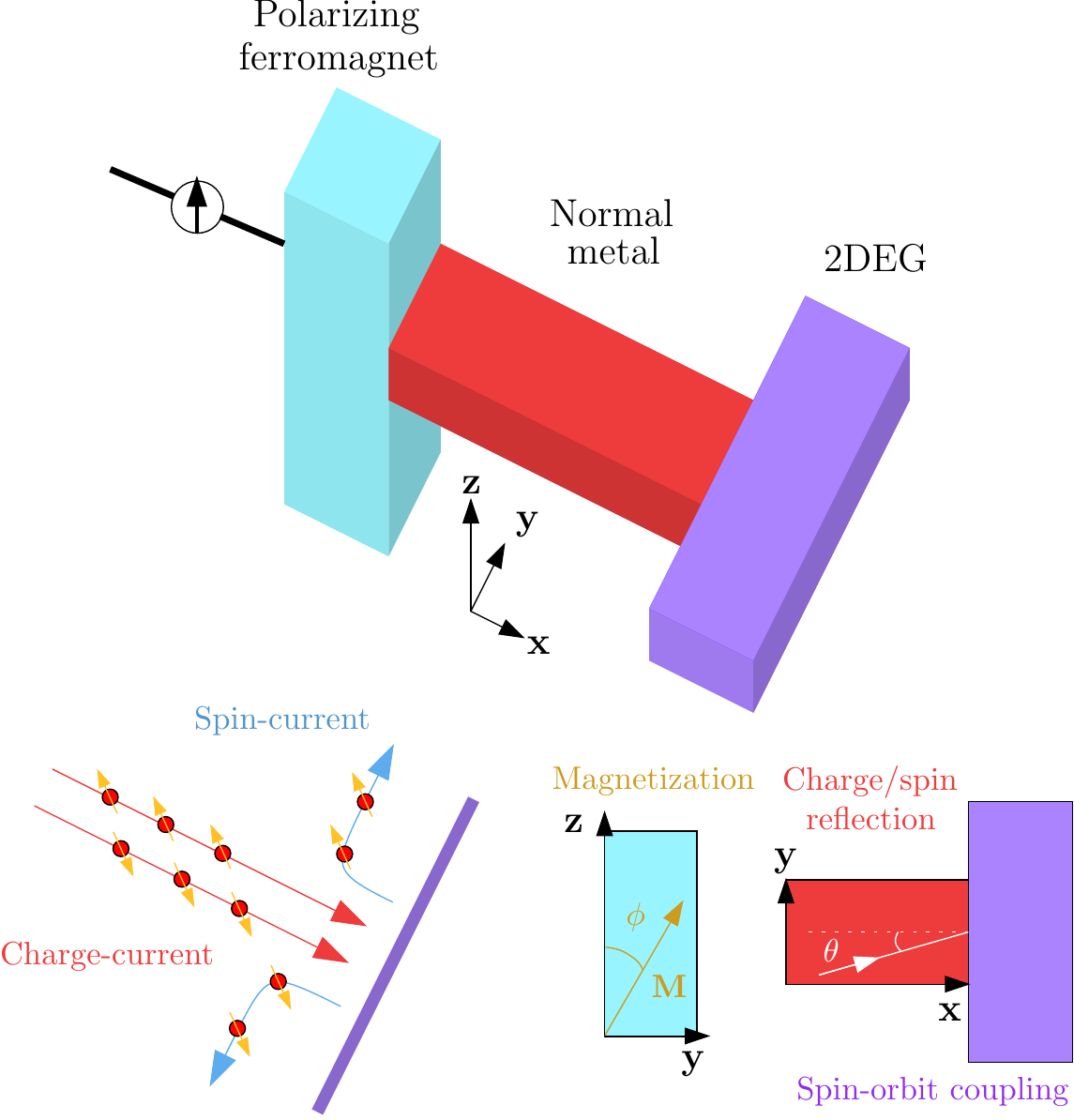}}
\caption{(Color online) A current bias is applied to a trilayer consisting of a polarizing ferromagnet, a normal metal, and finally a material with spin-orbit coupling, \eg a 2DEG. Due to reflection at the 
second interface bordering on the material with spin-orbit coupling, transverse charge- and spin-currents are induced in the normal metal region. The magnetization in the polarizing ferromagnet is misaligned an angle $\phi$ from the $\boldsymbol{z}$-axis, while the angle 
of incidence is denoted $\theta$. If the incident current is unpolarized, i.e. without the polarizing ferromagnet, a pure transverse spin-current is generated by means of spin-reflection off the barrier.  }
\label{fig:model}
\end{figure}

The influence of spin-orbit coupling on a spin-current is accompanied by welcomed as well as troublesome effects. On the one hand, the influence is beneficial in the sense that it offers a way of manipulating the spin-current of a system due to the coupling between the spin of the charge-carriers and an electric field. On the other hand, it is disadvantageous since it breaks conservation of spin and renders it a poor quantum number. It would be highly desirable to find a way of utilizing the first aspect of spin-orbit coupling and at the same time circumvent the difficulty associated with the latter. Here, we propose a way to achieve precisely this. 
\par
The experimental setup we have in mind is shown in Fig. 1. We assume that a charge-current, which may or may not be spin-polarized, flows into a normal metallic region that is sandwiched between the polarizing ferromagnet and a material with strong spin-orbit coupling, \eg a two-dimensional electron gas (2DEG). It should be noted that real ferromagnets do not act as perfect spin-polarizers, but it is nevertheless instructive to consider how the transport of charge and spin is influenced by a polarization of the incident current, similarly to Ref. \cite{stiles_prb_02} in the context of spin-transfer torque. As we shall see, the most interesting effects occur when the incident current is unpolarized, rendering the spin-polarizer obsolete. The chief motivation for including the polarizer is thus simply to gain a physical understanding of how the spin-polarization interacts with the spin-orbit coupling present in the 2DEG region.
\par
Although this setup is certainly simple, it offers some highly interesting possibilities with regard to the spin-currents flowing in the system. The crucial aspect is the scattering taking place at the interface between the normal region and the region with spin-orbit coupling. We will show how the scattering stemming from the spin-orbit coupling generates transverse currents flowing in the normal region. Since these currents flow in the normal region, they are \textit{not} subject to the difficulties associated with either the definition of the spin-current or the spin relaxation length hampered by spin-orbit coupling. Moreover, we find that these currents are highly sensitive to the spin orientation of the incoming current, i.e. the polarization direction $\phi$ in the ferromagnet. In fact, the transverse charge-current vanishes completely when $\phi=\pi/2$ for a Rashba-type spin-orbit coupling, regardless of the other parameters in the problem. The transverse spin-current, however, remains non-zero. Interestingly, we find that for both Rashba and Dresselhaus spin-orbit coupling, \textit{a pure transverse spin-current is generated in the normal metal when the incident current is completely unpolarized}, i.e. without the polarizing ferromagnet in Fig. \ref{fig:model}. This suggests that the charge- and spin-reflected currents induced in the normal metal region benefit from three major advantages: (i) the spin-current is conserved and its definition is unambiguous, (ii) the spin relaxation length is not influenced by spin-orbit coupling since it is absent in the normal region, and (iii) the charge- and spin-currents may be controlled in a well-defined way simply by adjusting the magnetization direction of the polarizing ferromagnet. Thus, the environment where the charge- and spin-currents of interest propagate (normal metal region) is non-hostile towards spin, while the control parameters tuning these currents are located in a different part of the system (the ferromagnet and the material with spin-orbit coupling) than where the actual currents propagate. This greatly facilitates the opportunity to exert control over the spin-current. It should be mentioned that in the context of mesoscopic spintronics, it is in general desirable and thus routine to consider leads without spin-orbit coupling in order to characterize spin transport in an unproblematic way. 

\begin{figure*}[t!]
\centering
\resizebox{1.0\textwidth}{!}{
\includegraphics{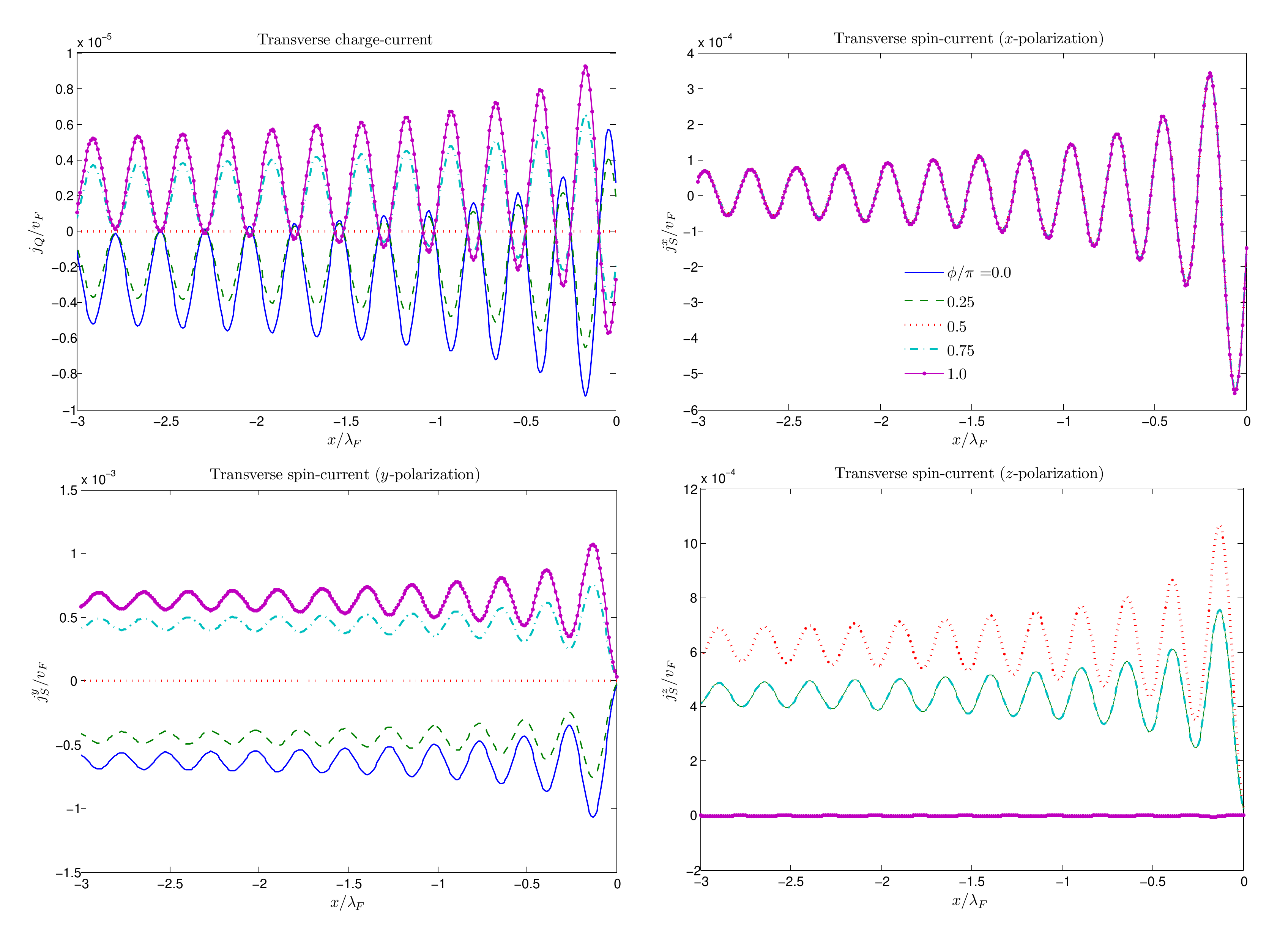}}
\caption{(Color online) Plot of the induced transverse charge- and spin-currents flowing parallel to the barrier in the normal metal region. We have set $\alpha=1\times10^{-4}$ and assumed a Rashba-type spin-orbit coupling. For $\phi/\pi=0.5$, the transverse charge-current vanishes, which makes it possible to obtain a pure spin-current signal by controlling the magnetization direction of the polarizing ferromagnet. The \textit{transverse charge-current also vanishes when the incident current is unpolarized}, i.e. a superposition of $\phi=0$ and $\phi=\pi$. This can be verified directly from the figure. Note that the curves for angles $\phi$ and $(\pi-\phi)$ are degenerate for the $x$- and $z$-polarizations of the spin-current.
}
\label{fig:spincurrent_rashba}
\end{figure*}

\section{Theory}
To address the above findings in a quantitative way, we employ a scattering matrix approach and calculate the resulting charge- and spin-currents in the system when a current bias is applied in the $\boldsymbol{x}$-direction. The spin-current is in general a tensor since it has a direction of flow in real-space and a polarization in spin-space. In the normal metal (N) region, we may write 
\begin{align}
\boldsymbol{j}_S = \text{Im}\{\psi^\dag \nabla \otimes \bs \psi\}/(2m_\text{N}),
\end{align}
where $\otimes$ is the tensor product between the gradient operator and the spin operator. In order to evaluate the spin-current, we need to construct the scattering states partaking in the transport processes. The quasiparticle states are obtained by solving the matrix equation which diagonalizes the Hamiltonian, namely:
\begin{align}\label{eq:bdg}
[\hat{H}_0(x) - h(x)\hat{\sigma}_z + \alpha(x)(k_y\hat{\sigma}_x - k_x\hat{\sigma}_y)] \psi = \varepsilon\psi,
\end{align}
with $\hat{H}_0 = [k^2/(2m(x)) -\mu]\hat{1}$. The effective electron mass $m(x)$ is assumed to be different in the normal metal and 2DEG regions. We have here taken into account the possibility of a magnetization in the region with spin-orbit coupling, assuming that it points along the $\boldsymbol{z}$-direction. The following derivations are made under the assumption of a  spin-orbit coupling of the Rashba-type [as employed in Eq. (\ref{eq:bdg})], but the procedure is identical for a Dresselhaus-type Hamiltonian where the spin-orbit coupling term reads $\alpha(k_y\hat{\sigma}_y - k_x\hat{\sigma}_x)$.
Similar Hamiltonians were considered also in Refs. \cite{linder_prb_07, mizuno_arxiv_09}. In order to gain some basic understanding of the role of spin-orbit coupling in our setup, we consider a N$\mid$2DEG junction with an incident spin-current at Fermi level from the N side. The interface is located at $x=0$, and hence $h(x) = h\Theta(x)$, $\alpha(x)=\alpha\Theta(x)$, with $\Theta(x)$ the Heaviside step function. The incident spin-current is assumed to be polarized in the $\boldsymbol{y}-\boldsymbol{z}$-plane with an angle $\phi$ relative to the $\boldsymbol{z}$-axis. Solving for the eigenvalues and eigenvectors of Eq. (\ref{eq:bdg}), we obtain the following wavefunctions:
\begin{align}\label{eq:waveN}
\psi_\text{N} = \Big[&\begin{pmatrix}
c\\
\i s\\
\end{pmatrix}\e{\i k_\theta x} + \Big\{r_\uparrow\begin{pmatrix}
c\\
\i s\\
\end{pmatrix}+ r_\downarrow\begin{pmatrix}
\i s\\
c\\
\end{pmatrix}\Big\}\e{-\i k_\theta x}\Big]\e{\i k_y y}
\end{align}
on the N side. It should be noted that considering both incident waves with $\phi=0$ and $\phi=\pi$ effectively gives an unpolarized incident current, which we shall comment on later. On the 2DEG side, we have:
\begin{align}\label{eq:psi2DEG}
\psi_\text{2DEG} = \Big[t_\uparrow \mathcal{N}_\uparrow\begin{pmatrix}
1\\
u_\uparrow\\
\end{pmatrix}\e{\i k_x^\uparrow x} + t_\downarrow \mathcal{N}_\downarrow\begin{pmatrix}
u_\downarrow\\
1\\
\end{pmatrix}\e{\i k_x^\downarrow x}\Big]\e{\i k_y y}
\end{align}
Above, we have defined $c=\cos(\phi/2)$, $s=\sin(\phi/2)$, and $k_\theta=k_F\cos\theta$, where $k_F=\sqrt{2m_\text{N}\mu_\text{N}}$ is the Fermi wavevector on the N side and $\theta$ is the angle of incidence. We have introduced the quantities 
\begin{align}
k_x^\sigma = \sqrt{(k^\sigma)^2 - k_F^2\sin^2\theta},
\end{align} 
in addition to $\mathcal{N}_\sigma = (1 + |u_\sigma^2|)^{-1/2}$ and
\begin{align}
k^\sigma &= [2m_\text{2DEG}\mu_\text{2DEG} + 2m_\text{2DEG}^2\alpha^2\notag\\
&+ 2m_\text{2DEG}\sigma\sqrt{h^2+m_\text{2DEG}^2\alpha^4 + 2m_\text{2DEG}\mu_\text{2DEG}\alpha^2}]^{1/2},\notag\\
&u_\sigma = -\frac{\sigma\alpha(k_y - \i\sigma k_x^\sigma)}{h + \sqrt{\alpha^2(k^\sigma)^2 + h^2}},\notag\\
\end{align}
All wave-functions are normalized to unity. For angles of incidence that satisfy $\sin\theta > k^\sigma/k_F$, the transmitted electrons become evanescent. To ensure their decay in the 2DEG region, one should then set $k_x^\sigma = \text{Re}\{k_x^\sigma\} + \i|\text{Im}\{k_x^\sigma\}|$.
The scattering coefficients $\{r_\sigma,t_\sigma\}$ are obtained by using the boundary conditions at the interface: \begin{align}
[r_m\partial_x&\psi_\text{2DEG}(x,y) - \partial_x\psi_\text{N}(x,y)]|_{x=0}= \hat{\eta}\psi_\text{N}(0,y),\notag\\
&\psi_\text{N}(0,y) = \psi_\text{2DEG}(0,y),\; \hat{\eta}=(2m_\text{N}V_0\hat{1} + m_\text{N}\alpha \i\hat{\sigma}_y).
\end{align}
Above, $r_m=m_\text{N}/m_\text{2DEG}$ denotes the ratio of the electron masses in the N and 2DEG regions. Employing the notation of Blonder-Tinkham-Klapwijk theory \cite{btk}, we define the dimensionless parameter $Z=2m_\text{N}V_0/k_F$ to characterize the interface transparency. The higher the value of $Z$, the stronger the interface barrier potential. An ideal interface is characterized by $Z=0$. The actual barrier potential is modeled as a delta-function and is proportional to $V_0$. Note that the boundary conditions properly take into account the off-diagonal elements in the velocity operator, as demanded in the presence of spin-orbit coupling \cite{molenkamp_prb_01}. 
The transverse charge-current $j_Q$ and spin-current $\boldsymbol{j}_S = (j_S^x,j_S^y,j_S^z)$ are finally obtained by integrating over all angles of incidence. Introducing a generalized current-vector $\boldsymbol{j} = (j_Q, \boldsymbol{j}_S)$ and $\hat{\boldsymbol{\tau}} = (\hat{1}, \hat{\boldsymbol{\sigma}}/2)$, we may write the transverse current as
\begin{align}\label{eq:current}
\boldsymbol{j} &= \int^{\pi/2}_{-\pi/2} \text{d}\theta\text{Im}\{\psi^\dag \partial_y \hat{\boldsymbol{\tau}} \psi\}/m_\text{N}.
\end{align}
Let us underline here that Eq. (\ref{eq:current}) naturally accounts for the contribution from different angles of incidence to the transverse current. In Eq. (\ref{eq:current}), the derivation operator $\partial_y$ brings a factor $\sin\theta$ to the integrand which thus ensures that angles of incidence close to $\pm\pi/2$ contribute strongly to the transverse current, as they should. If we had been concerned with the spin-current flowing perpendicular to the barrier, the replacement $\partial_y \to \partial_x$ would have been made, leading to a factor of $\cos\theta$ as usual in that case. In principle, one could also insert a weight-factor $f(\theta)$ inside the integrand of Eq. (\ref{eq:current}) which models a statistical distribution of the incoming particles. If the experimental geometry dictates that the incident quasiparticles are collimated near $\theta=0$, one could use \eg $f(\theta)=\cos\theta$. We have checked explicitly that our results undergo only a minor quantitative change when including such a weight-factor, and we here restrict our attention to the case without any such statistical distribution $f(\theta)$.

\begin{figure}[t!]
\centering
\resizebox{0.49\textwidth}{!}{
\includegraphics{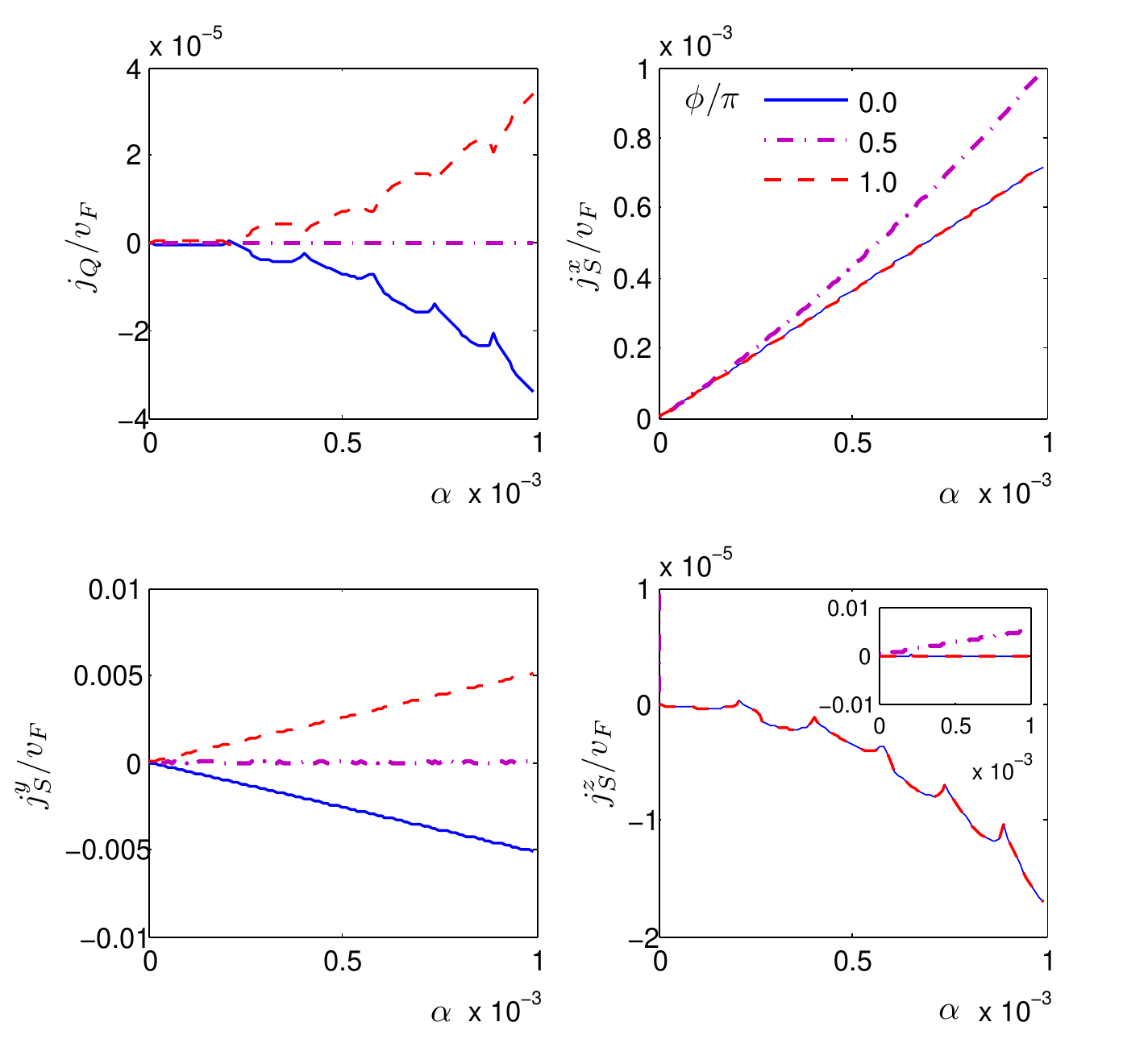}}
\caption{(Color online)  The transverse currents evaluated at $x/\lambda_F = -1.0$ as a function of the spin-orbit coupling strength $\alpha$ for Rashba-type spin-orbit coupling. In all cases, the magnitude of the current increases with $\alpha$.
}
\label{fig:strength_rashba}
\end{figure}

\section{Results}

Let us now discuss our choice of parameters for the physical quantities entering the model. Unless specifically stated otherwise, the figures are obtained using the parameter values below. We have distinguished between the electron masses and Fermi levels in the N and 2DEG region, as these differ greatly in realistic samples. In the normal metal region, we use $\mu_\text{N}=5$ eV and the electron mass $m_\text{N} = 0.51$ MeV. In the 2DEG region, we set $\mu_\text{2DEG} = 50$ meV with an effective electron mass $m_\text{2DEG}=0.1m_\text{N}$, i.e. $r_m = 10$. We set the spin-orbit coupling parameter to $\alpha = 1\times10^{-4}$ to model a typical value for a semiconductor \cite{gorkov_prl_01}. We have investigated numerically the influence of the exchange field in the 2DEG region, and found only minor quantitative changes in the results for values up to $h/\mu_\text{2DEG}=0.5$. Therefore, we shall here focus on the case of a 2DEG without magnetization, i.e. set $h=0$ and consider solely the effect of spin-orbit coupling. To model interface resistance, we set $Z=2m_\text{N}V_0/k_\text{F}=3$ as a reasonable measure for a rather low transmissivity interface. The angularly resolved transmission coefficient $T(\theta)$ is related to the barrier parameter $Z$ as $T(\theta) = 4\cos^2\theta/(4\cos^2\theta + Z^2)$. The Fermi-vector mismatch above may in principle be incorporated into a renormalized barrier potential $Z \to Z'$ with $Z'>Z$, thus lowering the transmissivity \cite{blonder_prb_83}.

\begin{figure*}[t!]
\centering
\resizebox{1.0\textwidth}{!}{
\includegraphics{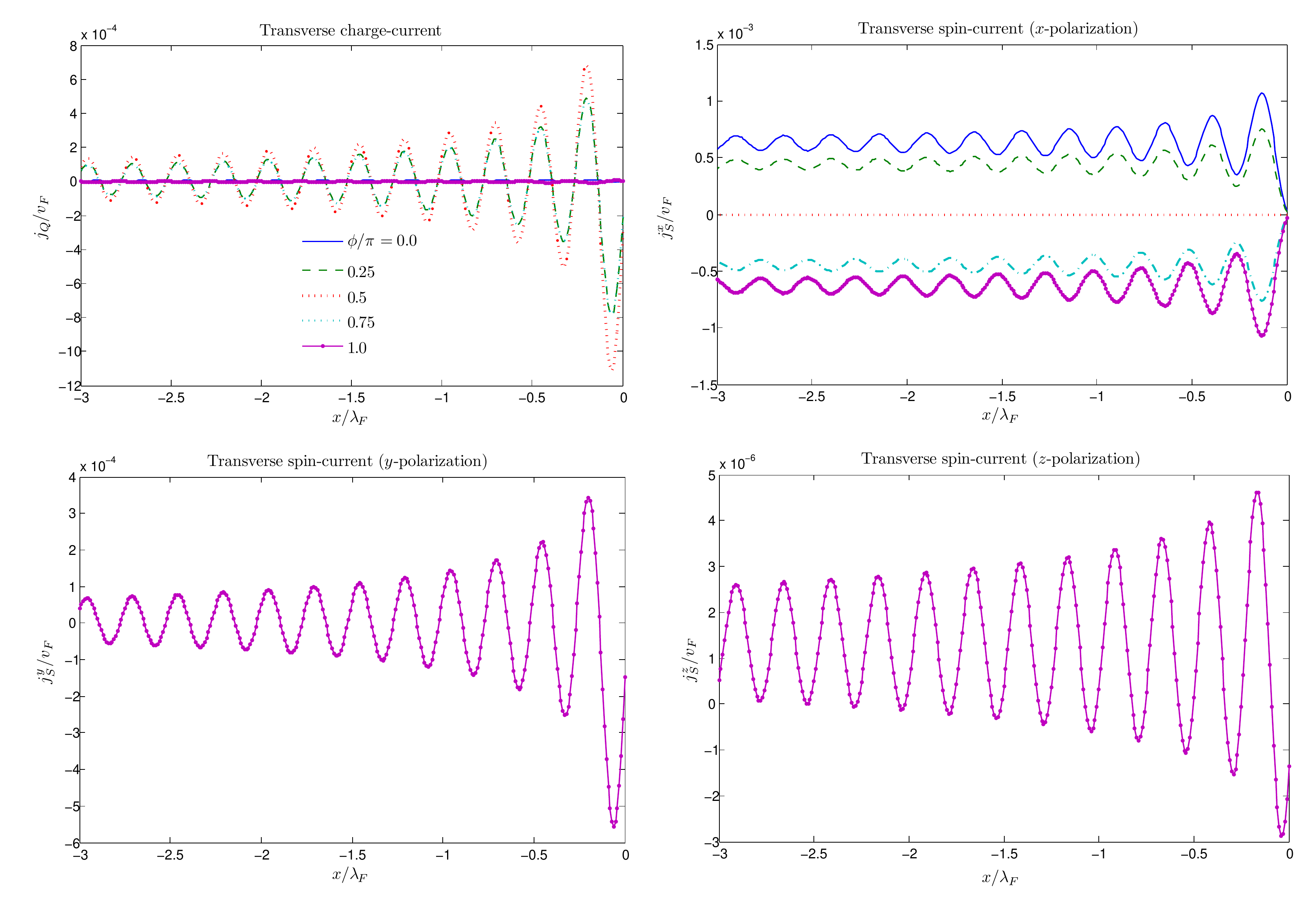}}
\caption{(Color online) Plot of the induced transverse charge- and spin-currents flowing parallel to the barrier in the normal metal region for a Dresselhaus-type spin-orbit coupling. The transverse charge-currents no longer vanishes at $\phi/\pi=0.5$, but for an unpolarized incident current (superposition of $\phi=0$ and $\phi=\pi$), the transverse charge-current is absent just as in the Rashba-case. 
}
\label{fig:spincurrent_dresselhaus}
\end{figure*}

\subsection{Rashba spin-orbit coupling}

We first investigate the transverse charge- and spin-currents flowing in the $\boldsymbol{y}$-direction when a current-bias is applied to the junction, using Eq. (\ref{eq:current}) with a Rashba-type spin-orbit coupling. The result is shown in Fig. \ref{fig:spincurrent_rashba} for several misorientation angles $\phi$ of the incident current. As seen, the presence of spin-orbit coupling induces non-zero transverse charge- and spin-currents in the normal region. The charge-current vanishes at $\phi/\pi=0.5$ \textit{or} if the incident current is completely unpolarized (i.e. a superposition of $\phi=0$ and $\phi=\pi$). This suggests a remarkable effect: simply by rotating the magnetization in the polarizing ferromagnet relative to the spin-orbit coupling vector that resides in the $xy$-plane, it is possible to tune the charge- and spin-currents in the normal metal region, and in particular one can obtain a \textit{pure spin-current signal} for $\phi/\pi=0.5$. Both the charge- and spin-currents display oscillations and decay to a constant, in general non-zero value in the bulk of the normal metal region. The oscillations appear as a result of interference terms of the type $\text{Re}\{r_\uparrow \e{-2\i k_\theta x}\}$ and $\text{Im}\{r_\downarrow \e{-2\i k_\theta x}\}$ generated when inserting the wavefunction Eq. (\ref{eq:waveN}) into Eq. (\ref{eq:current}). From these expressions, it is seen that as $|x|$ grows, the exponent varies more rapidly with $\theta$, such that the angular averaging in Eq. (\ref{eq:current}) eventually completely cancels out the $x$-dependent terms giving rise to the oscillations. The magnitude of the oscillations are therefore the strongest closest to the barrier $(x/\lambda_F\to0^-)$.
An important observation is that these transverse currents are not subject to the inherent problem with spin-orbit coupling related to the definition of the spin-current or spin-relaxation processes. We proceed to show that increasing spin-orbit coupling induces a stronger spin-reflected current, as depicted in Fig. \ref{fig:strength_rashba}. In all cases, the magnitude of the current increases with $\alpha$.

From Fig. \ref{fig:spincurrent_rashba}, we see that the charge current satisfies $j_Q(\phi) = -j_Q(\pi-\phi)$. The exact expressions for the reflection coefficients are too unwieldy to permit an analytical expression for $j_Q(\phi)$ through solving Eq. (\ref{eq:current}) by hand, but numerically we find that $j_Q(\phi) \propto \cos(\phi)$. As a result, it follows that for an incident current which is unpolarized, i.e. by removing the polarizing ferromagnet, a pure transverse spin-current may again be generated. We remind the reader that an unpolarized incident current can be thought of as a superposition of an incident $\phi=0$ and $\phi=\pi$ wave which leads to a total transverse charge-current:
\begin{align}
j_Q(0) + j_Q(\pi) = j_Q(0) - j_Q(\pi-\pi) = 0. 
\end{align}
The result is therefore a pure transverse spin-current. Effectively, this amounts to a conversion from a pure charge-current flowing in the $\hat{\boldsymbol{x}}$-direction to a pure spin-current flowing in the $\hat{\boldsymbol{y}}$-direction. Whereas such a scenario is also found inside a 2DEG subject to the spin-Hall effect, an important difference from our results is that in that case, the spin-current flows in the region where spin-orbit coupling effects are strong. Therefore, both the definition of the spin-current and its relaxation length become problematic. In our case, both of these issues are eluded since the spin-current flows in a normal metal region by means of reflection off a barrier in the presence of spin-orbit coupling. The vanishing of the transverse charge-current is understood by realizing that an injected unpolarized charge-current may be viewed as a coherent superposition of spin-$\uparrow$ and spin-$\downarrow$ electrons with equal weight. The two contributions are scattered in opposite directions due to the spin-orbit coupling,\cite{Govorov} and thus the net charge-current vanishes whereas the spin-current is non-zero.

\subsection{Dresselhaus spin-orbit coupling}

\begin{figure}[t!]
\centering
\resizebox{0.49\textwidth}{!}{
\includegraphics{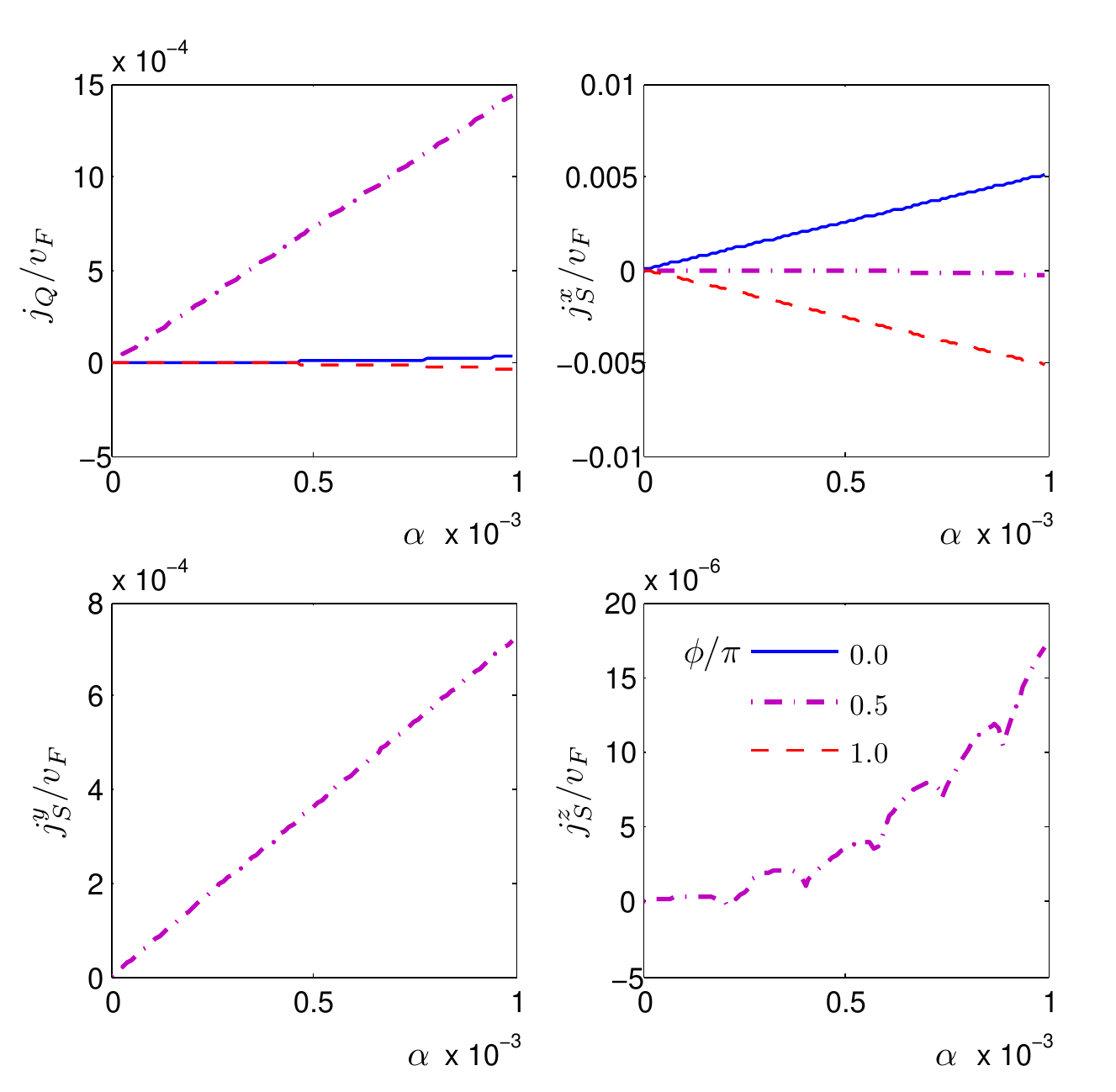}}
\caption{(Color online)  The transverse currents evaluated at $x/\lambda_F = -1.0$ as a function of the spin-orbit coupling strength $\alpha$ for Dresselhaus-type spin-orbit coupling. In all cases, the magnitude of the current increases with $\alpha$.
}
\label{fig:strength_dresselhaus}
\end{figure}

Let us also briefly investigate how the transverse charge- and spin-currents flowing in the $\boldsymbol{y}$-direction are influenced by a Dresselhaus-type spin-orbit coupling in the 2DEG region, in contrast to the Rashba-case treated in the previous section. In Fig. \ref{fig:spincurrent_dresselhaus}, we plot the transverse charge- and spin-currents flowing along the barrier. Due to the different structure of the Dresselhaus spin-orbit coupling compared to the Rashba-type, the $x$- and $y$-polarizations of the spin-current change roles. For the Dresselhaus-type, both the $y$- and $z$-polarization of the spin-current are insensitive to a variation in $\phi$. Similarly to the Rashba case, however, the magnitude of the transverse currents all increase with $\alpha$ as shown in Fig. \ref{fig:strength_dresselhaus}. It should be noted that the $x$-polarization of the spin-current does not vanish completely at $\phi/\pi=0.5$: it is simply strongly suppressed compared to the other values of $\phi$.

\section{Discussion}\label{sec:discussion}
The results presented in this paper suggest a method for obtaining a conversion from charge- to spin-currents. The method we propose 
exploits the possibility of manipulating the current by means of spin-orbit coupling, whereas it at the same time renders the spin-current 
immune towards the complicating and adverse effects of spin-orbit coupling with regard to relaxation processes and the very definition of 
a spin-current. 
In order to observe the transverse charge-spin current separation, one would need to find a way to probe the presence of a spin-flow in the 
transverse direction. This could in principle be achieved by measuring for instance spin accumulation at the edge of the normal metal wire with optical technique \cite{kato_science_04,wunderlich_prl_05}. 
\par
\begin{figure}[b!]
\centering
\resizebox{0.49\textwidth}{!}{
\includegraphics{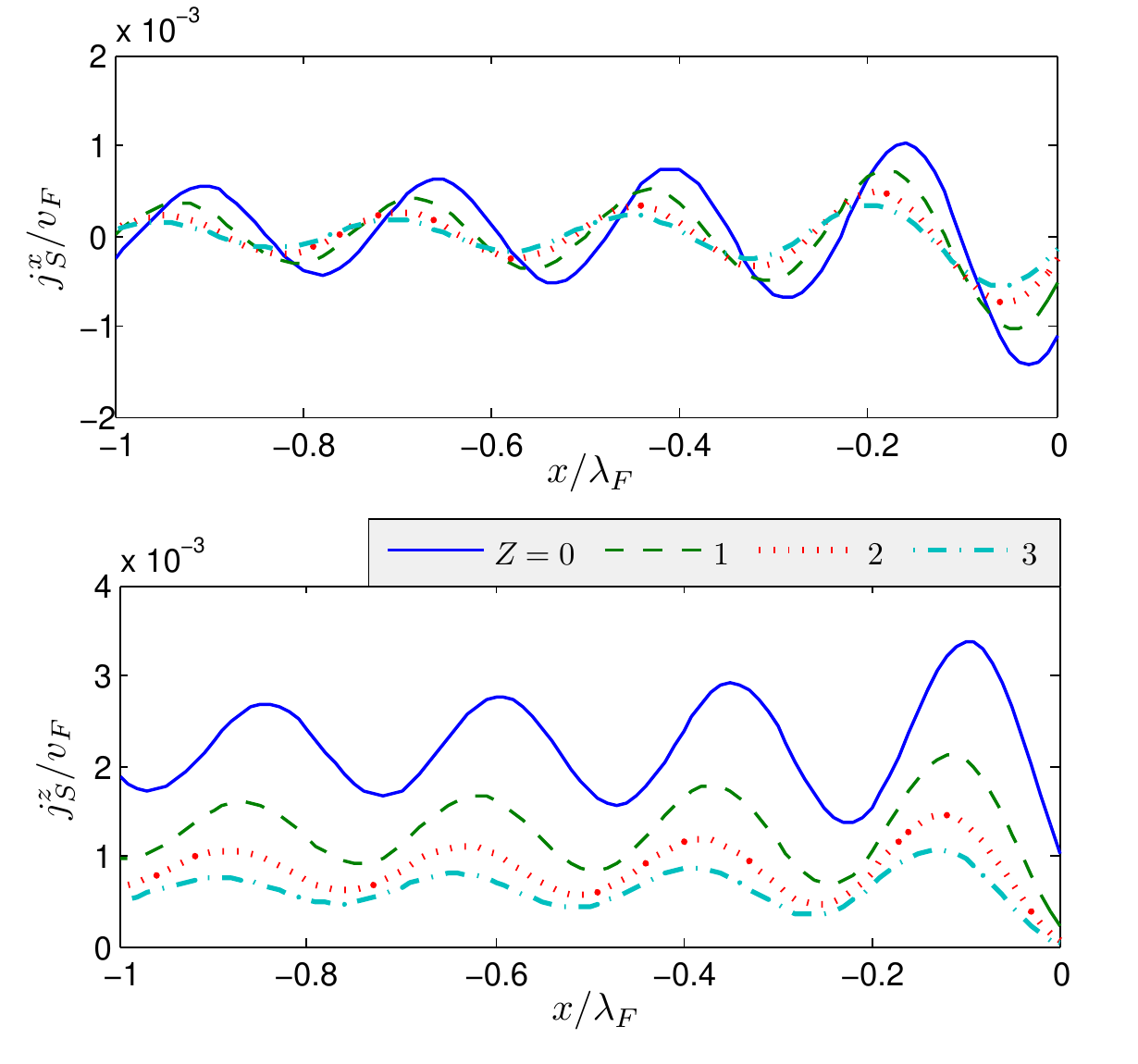}}
\caption{(Color online)  The transverse spin-currents $I_S^x$ and $I_S^z$ as a function of the spatial distance $x/\lambda_F$ penetrated into the normal metal region for Rashba-type spin-orbit coupling. We have here set $\phi/\pi=0.5$ and considered several barrier strengths $Z$.
}
\label{fig:barrier_rashba}
\end{figure}

Above, we kept the interface barrier potential fixed at $Z=3$, corresponding to a transmission coefficient of about $T\simeq0.3$ for normal incidence. To demonstrate that our results remain qualitatively unaltered upon varing the barrier potential $Z$, we plot in Fig. \ref{fig:barrier_rashba} the transverse spin-currents for an incident current with $\phi=\pi/2$. For a Rashba-type spin-orbit coupling, this is equivalent to the vanishing of transverse charge current, as seen in \eg Fig. \ref{fig:spincurrent_rashba}. The role of the barrier potential $Z$ is seen in Fig. \ref{fig:barrier_rashba} to simply reduce the magnitude of the spin-current, and does not influence the results qualitatively.

\begin{figure*}
\centering
\resizebox{1.03\textwidth}{!}{
\includegraphics{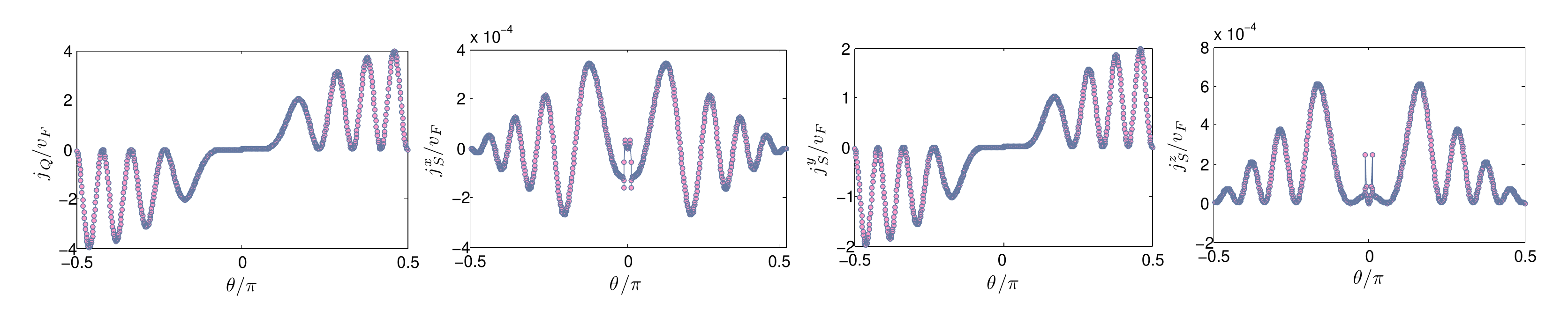}}
\caption{(Color online)  The angularly resolved transverse charge and spin-currents evaluated at $x/\lambda_F=-1.0$ in the normal metal region for a Rashba-type spin-orbit coupling. We have here set $\phi/\pi=0.5$.
}
\label{fig:angle_rashba}
\end{figure*}

It is also instructive to consider the angularly resolved transverse currents in order to understand the scattering processes on a more microscopic level. The transverse charge- and spin-currents are plotted in Fig. \ref{fig:angle_rashba} as a function of angle of incidence for a Rashba-type spin-orbit coupling, being evaluated at $x/\lambda_F=-1.0$. We here focus on the most interesting case $\phi/\pi=0.5$. As seen, both the charge-current and the $y$-polarization of the spin-current are antisymmetric around $\theta=0$, leading to a vanishing net current upon performing the angular integration. In contrast, the $x$- and $z$-polarizations of the spin-current are symmetric around $\theta=0$, yielding a net contribution to the total current. The specific form of the spin-orbit coupling potential should also influence the symmetry properties. In Fig. \ref{fig:angle_rashba}, we have used a standard Rashba-form with $\hat{H}_\text{SOC} = \alpha(k_y\hat{\sigma}_x - k_x\hat{\sigma}_y)$. In the case of a Dresselhaus-form $\hat{H}_\text{SOC} = \alpha(k_y\hat{\sigma}_y - k_x\hat{\sigma}_x)$, one would expect that the $x$- and $y$-polarization of the spin-current would interchange their symmetry properties since the two Hamiltonians are related by the substitution $\hat{\sigma}_x \leftrightarrow \hat{\sigma}_y$. This picture is verified by comparing the $x$- and $y$-polarizations of the spin-current in Fig. \ref{fig:spincurrent_rashba} with Fig. \ref{fig:spincurrent_dresselhaus}. The transverse charge-current nevertheless remains zero in both cases for an incident unpolarized current. We comment more on the role of adding a Dresselhaus term to $\hat{H}$ later in this section. Finally, we note that the oscillations of the currents in Fig. \ref{fig:angle_rashba} increase in rapidity as $|x|$ increases, i.e. farther inside the N region. The reason for this is the previously mentioned interference terms of the type $\text{Re}\{r_\uparrow \e{-2\i k_\theta x}\}$ and $\text{Im}\{r_\downarrow \e{-2\i k_\theta x}\}$ in the expression for the current.

\par
The assumption of ballistic transport and a sharp interface at the N$\mid$2DEG region is certainly an approximation to real systems where the 2DEG is often not characterized by the ballistic regime due to impurity scattering, \eg in InAs. Increased impurity scattering randomizes the momentum of scattered particles, which is detrimental to wavefunctions that are highly sensitive to the orientation of the momentum on the Fermi surface. Rashba spin-orbit coupling can be
interpreted as a wave vector-dependent Zeeman field that is altered dramatically when an electron scatters from one momentum orientation to another, even if the magnitude of the momentum remains the same. This type of scattering effectively randomizes the electron spin. 
\par
To show that the possibility of obtaining pure transverse spin-currents persists even in the nonballistic transport-regime, we consider a situation where both Rashba and Dresselhaus spin-orbit coupling are present and tuned to be of equal magnitude by means of proper gating, as discussed in Refs. \cite{schliemann_prl_03,Bernevig}. In this case, the scattering eigenvectors are independent of momentum, and thus survive angular averaging over the Fermi surface in the presence of non-magnetic impurities. To see this, consider the Hamiltonian
\begin{align}
\hat{H} = \hat{H}_0 + \alpha_R(k_y\hat{\sigma}_x -k_x\hat{\sigma}_y) +\alpha_D(k_y\hat{\sigma}_y - k_x\hat{\sigma}_x),
\end{align}
where $\alpha_R$ and $\alpha_D$ represent the spin-orbit coupling interaction parameter of Rashba and Dresselhaus type, respectively. In the case where these are equal, $\alpha_R=\alpha_D\equiv \alpha$, one obtains the eigenvalues
\begin{align}
\varepsilon_\pm = k^2/(2m)-\mu \pm \sqrt{2}\alpha(k_y-k_x),
\end{align}
with belonging eigenvectors
\begin{align}
\psi_+ = \frac{1}{\sqrt{2}}\begin{pmatrix}
\frac{1-i}{\sqrt{2}} \\
1 \\
\end{pmatrix},\; \psi_- = \frac{1}{\sqrt{2}}
\begin{pmatrix}
1 \\
-\frac{(1+\i)}{\sqrt{2}} \\
\end{pmatrix}.
\end{align}
The above wavefunctions $\psi_\pm$ are \textit{not sensitive to the direction of momentum}. In contrast, the wavefunctions in Eq. (\ref{eq:psi2DEG}) where a pure Rashba spin-orbit coupling was used are strongly dependent on the momentum orientation due to the $u_\sigma$ factors. In the present case of combined Rashba + Dresselhaus spin-orbit coupling, the transverse currents scattered off the barrier are obtained using a similar framework as described previously. Considering an injected unpolarized current, we find that the transverse charge-current again vanishes whereas the $x$- and $y$-components of the spin-current remain. Thus, the method of generating pure spin-currents suggested here should display robustness against impurity effects, although a more careful investigation of this matter certainly is warranted.
\par
Some previous works have also investigated spin-dependent scattering in hybrid structures in the presence of spin-orbit coupling \cite{khodas_prl_04, usaj_epl_05, palyi_prb_06, teodorescu_arxiv_09} as a possible mean of obtaining controllable spin-currents. In 
contrast to our results, however, the spin-currents obtained in these works suffer from all the problems 
related to spin-currents that we have elaborated on previously, since the spin-current flows in the 2DEG-region.  We emphasize that we 
have demonstrated the possibility of having a transverse, dissipationless spin-current \footnote{The spin-current is dissipationless in the sense that it is time-reversal symmetric.} in the absence of any accompanying charge-current. The charge-spin current separation could also find potential use as a spin-filter. Spin-filtering effects 
in spin-orbit coupled systems by selective angular beam injection have been discussed previously in Ref. \cite{tanaka_rapid_09}.
However, in previous discussions it has transpired that any net spin-current vanishes when taking into account all possible angles 
of incidence. In our case, the spin-current survives the averaging and is thus easier to access experimentally since angular filtering 
is a much more difficult task in quantum electronics than in, for instance, optics.

\section{Summary}

In summary, we have investigated the transport of charge and spin in a normal metal$\mid$2DEG junction, taking into account a spin-polarization of the incident current and a magnetic exchange energy in the 2DEG region. We find that it is possible to obtain a conversion from a pure charge-current to a pure spin-current simply by reflection off the barrier separating the normal and 2DEG region. More specifically, an incident unpolarized charge-current flowing towards the barrier is converted into a pure transverse spin-current flowing parallel to the barrier due to spin-dependent scattering off the barrier induced by the spin-orbit coupling. 
We emphasize that the spin-current flowing in the normal metal region is unambiguously defined and also rendered insensitive to the adverse spin-relaxation effects accompanying spin-orbit coupling. The method we propose to generate a pure spin-current in fact utilizes the desirable properties of spin-orbit coupling for facilitated control over spin-transport while simultaneously avoiding the complicating effects of spin-orbit coupling pertaining to the definition of the spin-current and spin-relaxation. Moreover, we have studied how the transverse charge- and spin-currents can be controlled by spin-polarizing the incident current. It is found that it is possible to tune the transverse charge-current to zero simply by rotating the magnetization of the polarizing ferromagnet, thus leaving a pure spin-current flowing parallel to the barrier. Our results may open up new perspectives for the generation and control over pure spin-currents.

\acknowledgments
J.L. and A.S. were supported by the Research Council of Norway, Grant No. 167498/V30 (STORFORSK). T.Y. acknowledges support by JSPS.

\appendix

\section{Spin-density continuity equation}

We here briefly outline the derivation of the continuity equation of the spin-density, setting $\hbar=c=1$. We consider the generic Hamiltonian 
\begin{align}
\hat{H} &= \frac{\vp^2}{2m} + \hat{H}_\text{SOC} + \hat{H}_\text{FM} + V(x),\notag\\
\hat{H}_\text{SOC} &= A(\vp)\hat{\sigma}_x - B(\vp)\hat{\sigma}_y,\; \hat{H}_\text{FM} = -\bs\cdot\h,
\end{align} 
with $V(x)$ containing all potential energy terms. Defining the spin-density as 
\begin{align}
\boldsymbol{S} = \psi^\dag \s \psi
\end{align} where $\s = \boldsymbol{\hat{\sigma}}/2$, we obtain 
\begin{align}
\i \partial_t \boldsymbol{S} = \i[\i (\hat{H}\psi)^\dag \s \psi - \i \psi^\dag\s \hat{H}\psi] = 2\i \text{Im}\{\psi^\dag \s \hat{H} \psi\}
\end{align}
by means of the Schr{\"o}dinger equation $\i\partial_t \psi = \hat{H}\psi$. Making use of the above equations, we obtain \begin{align}
\partial_t \boldsymbol{S} + \nabla \cdot \boldsymbol{j}_S = \mathcal{T}_\text{SOC} + \mathcal{T}_\text{FM},
\end{align}
where $\boldsymbol{j}_S$ is the conventional spin-current 
\begin{align}
\boldsymbol{j}_S = \text{Im}\{\psi^\dag \nabla \otimes \bs \psi\}/(2m).
\end{align}
Moreover, we have defined 
\begin{align}
\mathcal{T}_\text{SOC} = \text{Im}\{\psi^\dag \bs  \hat{H}_\text{SOC} \psi\},\; \mathcal{T}_\text{FM} = \text{Re}\{\psi^\dag (\bs \times \h) \psi\}.
\end{align}
In particular, the term $\mathcal{T}_\text{FM}$ may be interpreted as a spin-transfer torque to the magnetic order parameter, leading to a non-conserved spin-current even in the absence of spin-orbit coupling. The above treatment is valid both for Rasbha- and Dresselhaus-type spin-orbit coupling.

\end{document}